\documentclass[showkeys,showpacs]{revtex4}

\usepackage{graphicx}
\usepackage{dcolumn}
\usepackage{amsmath}
\makeatletter
\def\btt#1{\texttt{\@backslashchar#1}}
\DeclareRobustCommand\bblash{\btt{\@backslashchar}}
\makeatother

\begin{document}

\title[]{Gravitational collapse of null strange quark fluid and cosmic
 censorship}
\author{S. G. Ghosh}
\thanks{E-mail: sgghosh@iucaa.ernet.in}
\affiliation{%
Department of Mathematics, Science College, Congress Nagar,
Nagpur-440 012, INDIA}%

\author{Naresh Dadhich}
\thanks{E-mail: nkd@iucaa.ernet.in: To whom all correspondence to be done}
\affiliation{Inter-University Center for Astronomy and Astrophysics, \\
Post Bag 4, Ganeshkhind, Pune - 411 007, INDIA}%

\date{\today}
\begin{abstract}
We study gravitational collapse of the general
spherically symmetric null strange
quark fluid having the equation of state, $p = (\rho - 4B)/n$, where
$B$ is the bag constant.  An interesting feature that emerges is that the
initial data set giving rise to naked singularity in the Vaidya collapse of
null fluid gets covered due to the presence of strange quark matter component.
Its implication to the Cosmic Censorship Conjecture is discussed.
\end{abstract}


\maketitle

\section{Introduction}
A gravitating object when it undergoes indefinite collapse, the
end product is singularity which is marked by divergence of
physical parameters like energy density. As the singularity is
approached, density diverges and it would therefore be of relevance to
consider the
state of matter at ultra high density beyond the nuclear matter.
One of such possible states could be the strange quark matter which consists
of $u$, $d$ and $s$
quarks. It is energetically most favored state of baryon matter.
It could either be produced in quark-hadron phase transition in
the early Universe or at ultra high energy neutron stars
converting into strange (matter) stars \cite{ew}. In the context
of gravitational collapse, which is our concern here, it is the latter
process which would be pertinent.

The key question in the collapse process is whether the
singularity so formed will be visible or will be covered by an
event horizon prohibiting its visibility to an external observer.
That the latter is the case goes by the name of Cosmic Censorship
Conjecture (CCC) (see \cite{r1} for reviews of the CCC)), which
remains one of the most important unresolved issues in classical
general relativity. In the strong form of CCC, there could emerge
no null rays from the singularity in a reasonable space-time, and
hence it is invisible for all observers. That is, there occurs no
naked singularity for any observer. On the other hand the weak
form states that null rays can emerge from the singularity which
is however covered by an event horizon and hence they cannot reach
out the external observer. In the weak form singularity is locally
naked, say for an observer sitting on the collapsing star, but it
is globally not because it is safely hidden behind an event
horizon.

 There do however exist cases of regular initial data sets giving rise to
possibility of null rays emanating from the singularity and
reaching out to external observer \cite{pj}. That
 is, both the possibilities of singularity being naked and hidden inside a
black hole can occur. The critical question is what
decides between these two possibilities? In this context, it has
recently been argued that it is the shearing effect in the
collapsing inhomogeneous dust cloud that is responsible for the
ultimate outcome \cite{pnr}. This happens because shear produces distortion
in the collapsing fluid congruence which could cause distortion in the
geometry of the apparent horizon surface. Such a distortion of the apparent
horizon could let null rays emanating from the singularity to escape to
external observer. It turns out that if shear close to the center exceeds
a threshold limit, it gives rise to a naked singularity and else a
black hole.

Another critical question is, what is the state of matter as the
singularity is approached? It is certainly a case of diverging
density, and hence it would be appropriate to consider near the
singularity matter in the highest known density form. That brings
in the strange quark matter (SQM). Recently, collapse of
charged strange quark fluid together with the Vaidya null
radiation has been studied \cite{hc}. In this paper, we would
further like to analyze this process from the point of view of
formation of naked singularity and its strength, and more
importantly to bring out the effect of SQM. It turns out that
effect of SQM leads to covering up the region in the initial data
set window for naked singularity. That is, it tends to favor black
hole against naked singularity and consequently the CCC. This
happens because SQM contributes an additional attractive potential.

 The SQM fluid is characterized by the equation of state $p =
(\rho - 4B)/3$ where $B$ is the bag constant indicating
the difference between the energy density of the perturbative and
nonperturbative QCD vacuum, and $\rho, p$ are the energy density
and thermodynamic pressure of the quark matter \cite{ew,sw}. The fluid
consists of zero mass particles with the QCD
corrections for trace anomaly and perturbative interactions \cite{hc}. The
boundary of a strange star is defined by $p\to 0$ which would
imply $\rho\to B$. The typical value of the bag constant is of the
order of $B\approx 10^{15}g/cm^3$ while the energy density,
$\rho\approx 5 \times 10^{15} g/cm^3$ \cite{ew}. This shows that
SQM will always satisfy the energy conditions because $\rho\ge
p\ge 0$. We shall however consider the equation of state $p = (\rho - 4B)/n$
as a generalization of SQM fluid, and particularly the cases
$n = 2,\to\infty$ correspond to known cases of the Vaidya - de Sitter and the
Vaidya in constant potential bath collapse respectively.

In this paper, we
shall first obtain the general solution for SQM null fluid with the
generalized equation of state in the general spherically symmetric metric  in
the Bondi ingoing coordinates.  We shall then bring out explicitly the effect
of SQM on gravitational collapse in terms of covering of spectrum in the
initial data set for naked singularity by finding the threshold values for
the parameters involved in the  mass function.

 The paper is organized as follows. In the next section, we obtain the
general solution and analyze the collapse to show the effect of SQM in
shrinking the parameter window in the initial data set giving rise to
naked singularity. In the following section we discuss the
strength of the singularity and we conclude with a discussion.

\section{Strange quark null fluid collapse} \label{sec:gcsqm}
Expressed in terms of Eddington advanced time coordinate (ingoing coordinate)
 $v$,
the metric of general spherically symmetric space-time
 \cite{bi}
\begin{equation}
ds^2 = - A(v,r)^2 f(v,r)\;  dv^2
 +  2 A(v,r)\; dv\; dr + r^2 d \Omega^2 \label{eq:me2}
\end{equation}
$ d\Omega^2 = d \theta^2+ sin^2 \theta d \phi^2$.
Here $A$ is an arbitrary function.  It is useful to introduce
a local  mass function  $m(v,r)$  defined by $f = 1 - {2 m(v,r)}/{r}$.
For $m = m(v)$ and $f=1$, the
metric (\ref{eq:me2}) reduces to the standard Vaidya metric.
We wish to find the general solution of the Einstein equation for
the matter field given by Eq.(\ref{eq:emt}) for the metric
(\ref{eq:me2}), which contains two arbitrary functions. It is the field
equation $G^0_1 = 0$ that leads to $ A(v,r) = g(v)$. This
could be absorbed by writing $d \tilde{v} = g(v) dv$.  Hence,
without loss of generality, the metric (2) takes the form,
\begin{equation}
ds^2 = - \left[1 - \frac{2 m(v,r)}{r}\right] d v^2 + 2
d v d r + r^2 d {\Omega}^2 \label{eq:me}
\end{equation}
The energy momentum tensor for the null fluid together with SQM can be
written in the form \cite{hc,ww,vh}
\begin{equation}
T_{ab} = \mu l_a l_b + (\rho+p)(l_a n_b + l_b n_a) + p g_{ab}
\label{eq:emt}
\end{equation}
Here $\rho$ and $p$ are functions of $v$ and $r$ and
the two null vectors $l_{a}$ and $n_a$
\begin{eqnarray}
l_{a} = \delta_a^0, \: n_{a} = \frac{1}{2}
\left[ 1  - \frac{2 m(v,r)}{r} \right ] \delta_{a}^0 -
\delta_a^1
\nonumber \\
l^a = \delta_1^a,  \: n^a = - \delta_{0}^{a} -  \frac{1}{2}
\left[ 1  - \frac{2 m(v,r)}{r} \right ] \delta_{1}^a \nonumber
\\
l_{a}l^{a} = n_{a}n^{a} = 0 \; ~l_a n^a = -1.
\end{eqnarray}
Substituting this in Eq.~(\ref{eq:emt}), we find $ T^0_0 = T^1_1 = -\rho$,
$T^2_2 = T^3_3 = p$ and $T^1_0 = -\mu$, and the trace $T = -2 (\rho -p)$.
Here $\rho,~ p$ are the strange quark matter energy density and thermodynamic
pressure while $\mu$ is the energy density of the Vaidya null radiation.
The Einstein field equations now take the form
\begin{subequations}
\label{fe1}
\begin{eqnarray}
&& 4\pi\mu   = \frac{ \dot m}{r^2},   \label{equationa}
\\
&& 4\pi \rho = \frac{m ^{\prime } }{r^2},
\label{equationb} \\
&& 8 \pi p   =  -\frac{ m ^{\prime \prime}}{r}.
\label{equationc}
\end{eqnarray}
\end{subequations}
At higher density, the equation of state becomes uncertain as is the case
for nuclear matter and the strange quark matter would be no exception to it.
It is therefore appropriate to keep the coefficient $n$ free in the equation
of state,
\begin{equation}
p = \frac{1}{n}(\rho - 4B)  \label{eos}
\end{equation}
for SQM. We should however be open to the possibility that in the unknown new
super dense matter state, there could be altogether a different kind of
contribution which could entirely change the situation. So far SQM is the
most dense state of matter considered.

Imposing the equation of state and combining Eqs. ~(\ref{equationb}) and
(\ref{equationc}), we
obtain the following differential equation
\begin{equation}
m''(v,r) = - \frac{2}{nr} m'(v,r) + \frac{32 \pi B }{n} r  \label{de1}
\end{equation}
From this equation it is clear that the term involving the bag constant $B$
makes the contribution similar to the cosmological constant. We shall thus
seek the solution in the form,
\begin{equation}
m(v,r) = m_0(v,r) + \frac{\Lambda}{3}r^3  \label{m0}
\end{equation}
which would lead to
\begin{equation}
B = \frac{(n+1)\Lambda}{16\pi} \label{bc}
\end{equation}
and the differential equation
\begin{equation}
m_0''(v,r)  + \frac{2}{nr} m_0'(v,r) = 0   \label{de2}
\end{equation}
This has the general solution
\begin{equation}
m_0(v,r) = S(v)r^{(n-2)/n} + M(v)  \label{de3}
\end{equation}
The two arbitrary functions $M(v)$ and $S(v)$ would be restricted only by the
energy conditions. Here $\Lambda$ is not the cosmological constant but instead
is related to the bag constant $B$ via Eq.(\ref{bc}).
Thus the metric describing the radial collapse of null SQM in
$(v,r, \theta, \phi)$ coordinates reads as:
\begin{equation}
ds^2 =- (1  -  \frac{2 M(v)}{r} - \frac{2 S(v)} {r^{2/n}} -
\frac{\Lambda r^2}{3} ) dv^2  +  2 dv dr + r^2 d
\Omega^2 \label{eq:me1}
\end{equation}
This metric represents a
solution to the Einstein equations for a collapsing
null  SQM. The
physical quantities for this metric as in \cite{hc,ww} are given by
\begin{equation}
\mu = \frac{1}{4\pi r^2}
\left[\dot{M}(v)+\dot{S}(v) r^{(n-2)/n} \right]
\label{eq:mu}
\end{equation}
\begin{equation}
\rho = \frac{1}{4\pi r^2} \left[\frac{n-2}{n}S(v)r^{-2/n} + \Lambda
r^2 \right] \label{eq:rho}
\end{equation}
\begin{equation}
p = \frac{1}{4n \pi r^2} \left[\frac{n-2}{n}S(v)r^{-2/n} - \Lambda n
r^2 \right] \label{eq:p}
\end{equation}
Clearly, all the energy conditions would be satisfied for $n \geq 2$ because
it would ensure $\rho \geq0$ and $p \geq0 $, while $\mu \geq0$ would be taken
care of when we choose the mass functions for both null radiation and SQM. The
initial radius of the star from which the collapse begins would be given by
$p = 0$ which would also relate the bag constant with the mass function
$S(v)$.

 We are studying the collapse of SQM null fluid on a non-flat but empty
cavity. The
first shell arrives at $r=0$ at time $v=0$ and the final at $v=T$. A central
singularity of growing mass would develop at $r=0$.
 For $ v < 0$, $M(v)\;=\;S(v)\;=\;0$, i.e., we have
\begin{equation}
ds^2 = - (1 - \frac{\Lambda r^2}{3} ) dv^2 + 2 dv dr + r^2 d
\Omega^2 \label{eq:ime}
\end{equation}
 and for $ v > T$,
$\dot{M}(v)\;=\; \dot{S}(v)\;=\;0$, $M = M_0 > 0$. The space-time
for $v=0$ to $v=T$ is given by the generalized Vaidya metric
(\ref{eq:me1}), and for $v>T$ we have the generalized
Schwarzschild metric:
\begin{equation}
ds^2 = - (1  -  \frac{2 M_0}{r} - \frac{\Lambda r^2}{3} ) dv^2 + 2
dv dr + r^2 d \Omega^2 \label{eq:ome}
\end{equation}

\subsection{Occurrence of naked singularities} \label{second}
In this section, we adapt above solution to study the existence of
a naked singularity.
Let $K^{a} = {dx^a}/{dk}$ be the tangent vector to the null
geodesic, where $k$ is an affine parameter.   The geodesic
equations, on using the null condition $K^{a}K_{a} = 0$, take the
simple form
\begin{equation}
\frac{d^2v}{dk^2} + \frac{1}{r} \left[ \frac{M(v)}{r} + \frac{2
S(v)}{ r^{2/n}} - \frac{\Lambda}{3}r^2 \right]\left(\frac{dv}{dk}
\right)^2 = 0 \label{eq:kv1}
\end{equation}
\begin{equation}
\frac{d^2r}{dk^2} + \left[ \frac{\dot{M}(v)}{r} +
\frac{\dot{S}(v)}{r^{2/n}}
 \right] \left( \frac{dv}{dk} \right)^2 = 0 \label{eq:kr1}
\end{equation}
Radial ($ \theta$ and $ \phi \,=\,const$.) null
geodesics of the metric (11) must satisfy the null condition
\begin{equation}
\frac{dr}{dv} = \frac{1}{2} \left[1 -  \frac{2 M(v)}{r} -
\frac{2 S(v)}{r^{2/n}} - \Lambda \frac{r^2}{3} \right]
\label{eq:de1}
\end{equation}
Clearly, the above differential equation has a singularity at
$r=0$, $v=0$. If the singularity is naked, there must exist null ray
emanating from it. By investigating the
behavior of radial null geodesics near the singularity, it is therefore
possible to determine whether outgoing null curves meet the singularity in the
past. To go any further we would require specific form of functions $M(v)$
and $S(v)$, which we choose as follows:
\begin{subequations}
\label{FC}
\begin{eqnarray}
2 M(v)=&&\alpha v \;  (\alpha > 0),       \label{FCa}  \\
2 S(v)=&&\beta v^{2/n} \; (\beta > 0),  \; \label{Fcb}
\end{eqnarray}
\end{subequations}
Let $X \equiv v/r$ be the tangent to a possible outgoing geodesic
from the singularity. In order to determine the nature of the
limiting value of $X$ at $r=0$, $v=0$ on a singular geodesic, we
let $ X_{0} = \lim_{r \rightarrow 0 \; v\rightarrow 0} X =
\lim_{r\rightarrow 0 \; v\rightarrow 0} \frac{v}{r} $.
 Using
(\ref{eq:de1}), (\ref{FC})  and L'H\^{o}pital's rule we get
\begin{equation}
X_{0} = \lim_{r\rightarrow 0 \; v\rightarrow 0} X =
\lim_{r\rightarrow 0 \; v\rightarrow 0} \frac{v}{r}=
\lim_{r\rightarrow 0 \; v\rightarrow 0} \frac{dv}{dr} = \frac{2}{1
- \alpha X_0 - \beta X_0^{2/n} - \Lambda r^2/3} \label{eq:lm2}
\end{equation}
which implies,
\begin{equation}
\alpha X_0^2+  \beta X_0^{2/n+1} - X_0 + 2 = 0       \label{eq:ae}
\end{equation}
This is the equation which would ultimately decide the end state of collapse:
 a  black hole or a  naked singularity.

 Thus by analyzing
this algebraic equation, the nature of the singularity
can be determined.  The central shell focusing singularity would
atleast be locally naked (for brevity we have addressed it as naked
throughout this paper), if Eq.~(\ref{eq:ae})
 admits one or more positive real roots \cite{pj}.
 The values of the roots give the tangents of the escaping geodesics
near the singularity. When there are no positive real roots to
Eq.~(\ref{eq:ae}), there are no out going future directed null geodesics
emanating from the singularity.
Thus, the occurrence of positive roots would imply the violation
of the strong CCC, though  not necessarily of the weak form.
Hence in the absence of positive real roots,
the collapse will always lead to a black hole. The positive roots would
define the range for the tangent slopes from which the null geodesics can
escape to infinity. The critical slope would be given by the double root,
marking the threshold between black hole and naked singularity.

\subsubsection{Case $n=3$}
\begin{table}
\caption{Variation of $\alpha_C$ and $X_0$ for various
$\beta$ ($n=3$)} \label{table1}
\begin{ruledtabular}
\begin{tabular}{ccc}
$\beta$ & Critical Value$\alpha_C$ & Equal Roots  $X_0$ \\
\colrule
  0 & 1/8 & 3.9999 \\
  0.05 & 0.093728958525  & 4.18083  \\
  0.1 &  0.06294108366   & 4.39273  \\
  0.15 & 0.032689168719  & 4.64701  \\
  0.2 & 0.0030431168445  & 4.9625   \\
\end{tabular}
\end{ruledtabular}
\end{table}
We now examine the condition for the occurrence of a naked
singularity for $n=3$. First note that the Eq.~(\ref{eq:ae}) is
free of $\Lambda$ and hence it has no effect on the question under
study. However its presence makes the background space-time
asymptotically non flat. This happens because when $r \rightarrow
0$ the term ${\Lambda}r^2/3$ in Eq.(\ref{eq:lm2}) tends to zero.
For $\beta = 0$, the allowed range for $\alpha$ is given by (0,
1/8] as obtained earlier \cite{ns} for the Vaidya null radiation
collapse. In this case, it would be black hole for $\alpha > 1/8$.

\begin{table}
\caption{Values of Roots $X_0$ for $\alpha < \alpha_C$ for different
$\beta$ ($n=3$)} \label{table2}
\begin{ruledtabular}
\begin{tabular}{ccc}
  $\beta$ & $\alpha < \alpha_C$ & Roots $(X_0)$ \\
  \colrule
  0.0  & 0.1  & 1.40338, 1.9342  \\
  0.05 & 0.09 & 1.52319, 1.7237  \\
  0.1  & 0.06 & 1.55259, 1.74615 \\
  0.15 & 0.03 & 1.57943, 1.78409 \\
  0.2  & 0.003& 1.69165, 1.72026 \\
\end{tabular}
\end{ruledtabular}
\end{table}

 The numerical computation reveals that Eq.~(\ref{eq:ae}) would always admit
two positive roots for $\alpha \leq
\alpha_C$. Tangent to all outgoing radial null geodesics would lie in the range
$X_2 < X < X_1$, where $X_1$ and $X_2$ are the two roots. Table I shows the
critical values of $\alpha$ for various values of $\beta$. The window for
naked singularity is defined by (0,$\alpha_C$], and it is black hole for
$\alpha > \alpha_C$. Table II indicates
the slope range, given by the two roots, for the null geodesics to escape.
It is seen that $\alpha_C$ decreases with increase in $\beta$,
i.e., initial data set (0,1/8] for a naked singularity of the
Vaidya collapse shrinks by the introduction of SQM. There
exists a threshold value $\beta_T = 0.205198$ such that for $\beta
\geq \beta_T$, gravitational collapse of strange quark null fluid would
always end into a black hole for all $\alpha$.

Note that $\alpha$ refers to rate of collapse of the null radiation while
$\beta$ would refer to that of SQM. The $\beta$-threshold would therefore
define a critical rate of collapse for SQM required for collapse to end in a
black hole. Then it would fully respect CCC. Thus introduction of quark matter
favors formation of black hole.

\begin{table}
\caption{Variation of $\beta_T$ with $n$} \label{table3}
\begin{ruledtabular}
\begin{tabular}{cc}
  $2/n$ & $\beta_T$ \\
\colrule
  0.9     & 0.144    \\
  0.8     & 0.167    \\
  0.75    & 0.18     \\
  0.5     & 0.2052   \\
  0.25    & 0.45     \\
  0.125   & 0.6194   \\
  0.0625  & 0.756    \\
\end{tabular}
\end{ruledtabular}
\end{table}

\begin{table}
\caption{Values of equal roots $X_0$ for different $n$
and  $\beta$} \label{table4}
\begin{ruledtabular}
\begin{tabular}{cccc}
 $2/n$ &  $\beta$ &   $ \alpha_C$ & Equal Root $(X_0)$ \\
\colrule
  0.90   &0.14 &0.00343411423125 & 4.21544  \\
  0.80   &0.16 &0.00502531901422 & 4.47484  \\
  0.75   &0.17 &0.00679411554354 & 4.61836  \\
  0.5    &0.20 &0.0306384999095  & 5.1784   \\
  0.25   &0.44 &0.00178952699837 & 9.51405  \\
  0.125  &0.61 &0.00077533872957 & 16.519   \\
  0.0625 &0.74 &0.00061702385153 & 27.1684  \\

\end{tabular}
\end{ruledtabular}
\end{table}

\subsubsection{Other Cases}
\paragraph{} Case $n =2$\\
Then we find that
\begin{equation}
\mu = \frac{1}{4 \pi r^2} \left[ \dot{M}(v) + \beta \right],  \hspace{.2in}
\rho = - p = \frac{\Lambda}{4 \pi}
\end{equation}
and the algebraic equation takes the form
\begin{equation}
(\alpha +  \beta) X_0^2 - X_0 + 2 = 0       \label{eq:ae1}
\end{equation}
The metric in this case takes the form of the
Vaidya - de Sitter metric.  The singularity
is visible for $(\alpha +  \beta) < 1/8$. It is the null fluid collapse in
the background of the de Sitter space, where $\Lambda$ is generated by the
bag constant.
\paragraph{} Case $n\rightarrow\infty$. \\
We have
\begin{equation}
\mu =  \frac{1}{4\pi r^2}  \dot{M}(v),  \hspace{.2in} p=0, \hspace{.2in}
\rho = \frac{B}{4\pi r^2} \label{eq:ni}
\end{equation}
In this case we have the dual Vaidya metric or Vaidya metric with constant
potential \cite{jdj}.  The algebraic equation:
 $\alpha X_0^2 + (\beta - 1)  X_0 + 2 = 0$,
would admit a positive root for $\alpha \leq 1/8 (\beta-1)^2 $, giving the
range for naked singularity as obtained in \cite{jdj}. This is simply the
null fluid collapse in the background of constant potential which is
characterized by $T^0_0 = T^1_1 = const./r^2$, as is the case in
Eq.(\ref{eq:ni})
above. Note that $\beta < 1$ else the metric signature would change.
\paragraph{} Other $n$ \\
We also note that as $n$ increases, so does the threshold value $\beta_T$.
This is shown in Tables III and IV.

\section{Curvature Strength of Singularities}
An important aspect of a singularity is its gravitational strength
\cite{ft1}.  A singularity is gravitationally strong in the sense
of Tipler, if it destroys any object which falls into it and weak
otherwise.  It is now widely believed that space-time does not
admit an extension through a strong curvature singularity, i.e.,
space-time is geodesically incomplete. Through a weak singularity,
space-time could be analytically extended to make it geodesically
complete. There have been attempts to relate strength of a
singularity to its stability  \cite{djd}. Recently, Nolan
\cite{bc} gave an alternative approach to check the nature of
singularities without having to integrate the geodesics equations.
It was shown \cite{bc} that a radial null geodesic which runs into
$r=0$ terminates in a gravitationally weak singularity if and only
if $dr/dk$ is finite in the limit as the singularity is
approached (this occurs at $k=0$, the over-dot here indicates
differentiation along the geodesic). If the singularity is weak,
we have
\begin{equation}
\frac{dr}{dk} \sim  d_{0} \hspace{0.2in} r \sim  d_{0} k
\end{equation}
Using the asymptotic relationship ($dv/dk \sim d_0 X_0$ and
 $v \sim d_0 X_0 k  $) and Eq. (\ref{FC}), the
geodesic equation yields
\begin{equation}
\frac{d^2v}{dk^2} \sim  - (\alpha X_0 d_0^{-1}k^{-1} + \beta X_{0}^{2/n}
d_{0}^{-1} k^{-1} - \frac{\Lambda}{3}d_{0}k)d_0^2 X_0^2
\end{equation}
But this gives
\begin{equation}
\frac{d^2v}{dk^2}  \sim c k^{-1}
\end{equation}
where $c = (\alpha X_0^{(n-2)/n} + \beta)X_0^{2(n+1)/n} d_0^{-1}$.
This is inconsistent
 with $dv/dk \sim  d_{0} X_{0}$, which is finite.  Since the coefficient
$c$ is non-zero, the naked singularity is
gravitationally strong in the sense of Tipler \cite{ft}. Having
seen that the naked singularity is a strong
curvature singularity, we check it for scalar polynomial
singularity. The Kretschmann scalar for the metric (\ref{eq:me1}) 
with the prescriptions
(\ref{FC}), takes the form
%\begin{widetext}
\begin{eqnarray}
K = && \frac{4}{3n^4 r^4} \Bigg[ \alpha^2 n^4 X_0^2 + (4 \beta
\Lambda n^2 -6 \beta  \Lambda n^3 + 2  \beta \Lambda n^4)
X_{0}^{2/n} +(12 \alpha \beta \Lambda n^2 + \nonumber \\
&& 8  \alpha \beta \Lambda n^3+6 \alpha \beta \Lambda n^4)
X_{0}^{(n+2)/2}
  \nonumber \\
+&& (12 \beta^2 n + 15 \beta^2 n^2 +
 3 \beta^2 n^4) X^{4/n}  \Bigg]
+ \frac{8}{3} \Lambda^2 \label{eq:ks1}
\end{eqnarray}
%\end{widetext}
which diverges at the naked singularity and hence the singularity
is also a scalar polynomial singularity. The Ricci scalar also
diverges. It however vanishes for the  Vaidya space-time
\cite{bs}. Thus the naked singularities studied here are strong
curvature singularity and hence are physically significant.
Lastly, we shall calculate Weyl scalar
\begin{equation}
C = \frac{4}{3 n^4 r^4} \left[\alpha n^2 X+2 \beta X^{2/n}+3 \beta X^{2/n}
+\beta n^2 X^{2/n} \right]^2
\end{equation}
which too would diverge. The Weyl curvature describes non local effects of
gravitation produced by free part of the field. It is generated by
inhomogeneity and anisotropy, particularly divergence of shear \cite{ellis}.
In the context of naked singularity, like shear and inhomogeneity the Weyl
curvature would also play significant role.

\section{Discussion}
In this paper, we have obtained the general solution
 for null SQM fluid with the equation of
state given by Eq.~(\ref{eos}) for the general spherically
 symmetric metric ~(\ref{eq:me}) in
the Eddington advanced time coordinate (ingoing coordinates).
 We have used the solution to study the end state of the
collapse. The present case is an example of
non self-similarity as well as non asymptotic flatness, and yet
there does occur a regular initial data set which would lead to
naked singularity. 

The relevant question is what effect does the presence of the SQM have on
formation or otherwise of a  naked singularity. Our results imply that
the presence of SQM leads to shrinking of the initial data space for
 naked singularity of the Vaidya null fluid collapse. That is, it tends to
favor black hole. This tendency is caused by the additional attractive
potential, varying as $r^{-2/n}$, produced by SQM which results in
strengthening of gravity. There also exists a threshold value $\beta_T =
0.205198$ such that for
$\beta  \geq \beta_T$ the end state of collapse of null
SQM is always a black hole for all $\alpha$. This is the critical value of
the rate of collapse of SQM for respecting CCC. The case $n=3$ represents the
zero mass particles with QCD corrections for trace anomaly and perturbative
interactions \cite{hc}.
The energy conditions require $n\geq2$. The case of $n=2$
corresponds to the null fluid collapse in the background of de Sitter space
where the bag constant provides the $\Lambda$. The case $n\rightarrow\infty$,
corresponds to the null fluid collapse in the background of constant potential
space as studied in \cite{jdj}. These are the two extreme limiting cases
encompassing the physically allowed cases. Though there is no much physical
motivation in the context of SQM for $n\neq3$ cases, yet these two particular
cases are interesting. That is at least these three cases could be considered
in a unified equation of state given by Eq. (8).

 As mentioned earlier the strong version of the CCC doesn't allow even
 locally naked singularity, i.e., the space-time should be globally
hyperbolic.  It turns out that necessary and sufficient condition for
a singularity to be locally naked is that the algebraic Eq.~(\ref{eq:ae})
should have atleast one or more positive root \cite{pj}.
Hence existence of the positive roots of Eq.~(\ref{eq:ae}) is a
counter example to the strong version of the CCC.
In the absence of the proof of any version of the CCC, such examples remain
the only tool to study this important and unresolved problem.

Quark stars could be formed in the realistic astrophysical setting. The
core collapse of a massive star after the supernova explosion sets
in first and second order phase transitions which result into
deconfined quark matter. The
 other possibility is that some neutron stars could accrete matter and
undergo phase transition to turn into quark stars
\cite{cdl,dpl,cd}. Thus study of gravitational collapse with quark matter
component is quite in order because it is perhaps astrophysically more
realistic. In the ultimate stage of collapse close to the singularity,
density is diverging. The quark matter contribution would therefore perhaps
be most significant in deciding the ultimate result of the collapse.

\acknowledgements 
SGG would like to thank IUCAA, Pune for hospitality
 while this work was done and  UGC, Pune (INDIA) for Grant
No. MRP F. No 23-118/2000 (WRO).  The invariants in section III
has been calculated using GRTensorM \cite{grt}.
Authors are gratefult to the referee for constructive criticsm.

\noindent

\end{document}